# Piezoelectric MEMS Phase Modulator for Silicon Nitride Platform in the Visible Spectrum

Firehun T. Dullo[*,1], Paul C. Thrane[1], Nikhil Jayakumar[2], Zeljko Skokic[1], Christopher A. Dirdal[1], Jo Gjessing[1] and Balpreet S. Ahluwalia[2,3]

[1]Department of Microsystems and Nanotechnology, SINTEF Digital, Gaustadalleen 23C, 0373 Oslo, Norway
[2]Department of Physics and Technology, UiT- The Arctic University of Norway, Klokkargårdbakken 35, 9037 Tromsø, Norway
[3]Department of Physics, University of Oslo, Oslo, Norway

*E-mail: firehun.t.dullo@sintef.no

## Abstract

Active photonic integrated circuits (PICs) in the visible spectrum are essential for on-chip applications, requiring low-loss waveguides with broad transparency and efficient, low-power phase modulation. Here, we demonstrate a compact, ultra-low-power phase modulator based on silicon nitride ($Si_3N_4$) waveguide integrated with thin-film lead zirconate titanate (PZT) that actuates a bridge-type MEMS. The suspended actuator exploits PZT's strong piezoelectric effect to induce mechanically driven phase shifts, enabling efficient modulation in a Mach–Zehnder interferometer. For 3 mm and 5 mm modulators, phase shift of 1.45 π and 2.5 π are achieved at 10 V, corresponding to a scalability metric (Vπ·L) of 2.25 V·cm at 635 nm. This is an order-of-magnitude improvement in scalability over stress-optic PZT modulators. The devices also exhibit ultralow power consumption (~12 nW), ~ 5 ms rise time, and optical loss < 0.75 dB/cm. Furthermore, we showcase on-chip beam shaping.

# Introduction

Photonic integrated circuit (PIC) technology is having a transformational effect on conventional free-space optics by integrating multiple optical components with diverse functionalities into a single compact chip.[1] Early PICs were predominantly passive, enabling static optical routing, splitting and filtering of light. More recently, tunable (active) and reconfigurable PICs have emerged, providing dynamic control of guided optical modes for modulation, switching and amplification. Such active control is essential for a wide range of on-chip photonics applications, including optical phased array[2], photonic neural networks,[3] programmable PICs,[4] biomedical sensing,[5] bio-imaging,[6] quantum technologies[7] and light detection and ranging (LiDAR).[8]

At the core of active PICs lie tunable elements that operate through mechanisms such as electro-optic modulation, thermal tuning, microelectromechanical systems (MEMS) actuation and stress-optic effects.[9] Tunable elements with these mechanisms enable key active photonics components, including phase and amplitude modulators, optical switches and tunable couplers. Active PICs are well established in the infrared (IR) regime, primarily using silicon (Si) via carrier injection or depletion, and indium phosphide (InP) through the electro-optic effect.[10,11] However, both Si and InP are limited to IR operation due to their material transmission windows. Lithium niobate ($LiNbO_3$) has recently emerged as a leading material for broadband active PICs, offering a strong intrinsic electro-optic response and a wide transparency window extending from the visible to IR.[12] In contrast, silicon nitride ($Si_3N_4$) remains the most widely used platform for visible-spectrum PICs due to its mature fabrication process, low propagation loss, broad transparency (visible to IR), high power-handling capacity, and CMOS compatibility. However, unlike $LiNbO_3$, $Si_3N_4$ is inherently passive with poor thermo-optic, electro-optic and piezoelectric coefficients. To introduce active functionality, $Si_3N_4$-based PICs must therefore incorporate additional thin film materials with suitable tunable properties.

The focus of this work is development of novel active PIC using the $Si_3N_4$ platform for visible spectrum. Therefore, we summarize different phase modulator techniques explored so far for $Si_3N_4$-based PICs operating at visible wavelengths. These can be broadly classified as thermo-optic modulation, Pockels electro-optic (EO) modulation, stress-optic modulation and MEMS actuator-based modulators.

The most-widely reported integrated phase shifters for $Si_3N_4$ operating in the visible are based on the thermo-optic effect. Thermo-optic modulation, which changes the refractive index through localized heating, has been implemented in $Si_3N_4$ PICs by integrating thin films such silicon dioxide, polymers and non-volatile phase change materials[13]. Improvements in the efficiency of this technique have focused on (1) enhancing thermal isolation using suspended structures: under cuts and trenches[14], and (2) using advanced polymers with high thermo-optic properties[15]. While these designs show reduced power consumption and footprint of the active element, they often introduce trade-offs in optical loss, bandwidth, and fabrication complexity. The state-of-the art $Si_3N_4$ thermo-optic phase shifter[13] has a power consumption of around 1.2mW at 561 nm with an active element length of 1.5 mm and insertion loss of 5.5dB/cm.

Electro-optic (Pockels) modulators, which exploit electric-field induced refractive index changes, have been realized in $Si_3N_4$ PICs operating in the visible spectrum by integrating thin-film materials such as ferroelectric lead zirconate titanate PZT[16], $LiNbO_3$[17] and zinc oxide (ZnO) or zinc sulfide (ZnS) thin films[18]. More recently, electric field induced charge displacement in $Si_3N_4$ material has been shown to induce a second-order nonlinearity, enabling a monolithic linear electro-optic effect[19]. Pockels-based modulators offer fast, power-efficient operation but introduce elevated optical losses into the waveguide.

Stress-optic modulation, in which the refractive index changes due to mechanically induced stress, has been demonstrated for $Si_3N_4$ PICs operating in the visible spectrum using integrated PZT thin films [20–22]. Hosseini *et al.* demonstrated such a modulator on the TriPlex platform on $Si_3N_4$ at 640 nm[21] and Everhardt et al., subsequently optimized the design with a dome-structured PZT layer, achieving half - wave-voltage-length product (Vπ·cm) of 16 V·cm and a voltage length optical loss product (Vπ ·L·α) is 1.6 V·dB[23]. This approach offers low-power operation and moderate speed but typically requires longer active element footprints.

MEMS-based modulation, which relies on mechanical deformation of the waveguide to alter the optical path length and refractive index change due to the mechanical induced stress, has been demonstrated in visible spectrum $Si_3N_4$ PICs using electrostatic MEMS[24] and piezoelectric MEMS[25] actuators based on aluminum nitride (AlN). Piezoelectric MEMS actuators based on PZT has so far been demonstrated only in the near-IR[26]. The superior piezoelectric response of PZT compared to electrostatic and piezoelectric responses of AlN for the applied voltage[27] suggests strong potential for visible-wavelength operation. The state-of-the-art AlN based piezoelectric MEMS for $Si_3N_4$ waveguide reported so far demonstrated scalability metric of 6 V.cm and power consumption below 30 nW.

The most explored techniques in the visible range rely on refractive-index modulations (dn). However, even when long modulators are used, the achievable dn is typically very small (on the order of $10^{-5}$), resulting in only small phase shifts (≈ dn·L), where L is the modulator length. In contrast, directly modulating the optical path length of the waveguide via modulator (dL) by even a few hundred nanometers can generate a much larger phase shift (≈ $n_{eff}$·dL), where $n_{eff}$ is the effective index of the guided mode (typically between 1.5 to 2 for waveguides based on high index contrast materials). Advancing PZT-MEMS-based visible-wavelength modulators thus represents a critical step toward extending the functionality and scalability of $Si_3N_4$ photonics platforms for next-generation active PICs. In this work, we present a PZT-MEMS-based visible-wavelength phase modulator realized by integrating a PZT stack onto a cladded $Si_3N_4$ waveguide and employing a suspended bridge-like actuator produced through selective front and back side etching as outlined in Fig. 1 Applying an electric field across the PZT layer induces in-plane contraction and out-of-plane expansion, mechanically deforming the suspended PZT stack along with underlying clamped waveguide. This mechanical deformation of the optical waveguide changes the optical path length of the guided mode in the waveguide. We demonstrate the phase modulator integrated into a Mach-Zehnder interferometer (MZI) as a phase modulator (Fig. 1 a) using suspended $Si_3N_4$ waveguide geometry and showcase its utility for on-chip beam shaping.

By leveraging the suspended $Si_3N_4$ waveguide geometry and the superior piezoelectric coefficient of PZT ($e_{31,f}$ = -16 $C/m^2$)[27], the proposed design achieves a large phase shift per applied voltage, significantly reducing the actuator footprint compared to the stress-optic based phase modulator[23]. For 3- and 5-mm active elements, we demonstrate a 1.4π and 2.5π phase shift at 10V, respectively, corresponding to a scalability metric, Vπ·L of 2.25 V·cm. The actuator exhibits ultralow power consumption (12 nW) and a measured optical loss below 0. 75V.dB, confirming the low-loss performance. The proposed phase modulator offers an order of magnitude scalability improvement over the state-of-the-art stress optic PZT modulators (16 V·cm)[23] and a 50% improvement compared to the AlN based piezoMEMS phase modulator (6 V·cm)[25].

Being a mechanical system, the modulation speed of the PZT-MEMS is from Hz to MHz which is inferior to the electro-optic modulators (GHz). However, for on-chip microscopy targeting life sciences application, the camera acquisition time (typically 0.1-100 ms) eventually limits the modulation speed requirement. Thus, modulation speed in a range of <1 KHz is more than sufficient for on-chip microscopy and sensing application in the visible range. Such on-chip beam engineering via phase shifters has been explored for bioimaging applications, using on-chip structured illumination microscopy (SIM)[6] methodology on $Si_3N_4$ platform. On-chip SIM requires precise and reproducible phase shifting between the evanescent interference fringes to generate a super-resolved image. The fringe movement for visible spectra is typically accomplished using thermo-optical phase modulators, these require high power consumption and exhibit significant heat crosstalk between the phase shifters. In this work, we implemented an interferometer design to generate evanescent interference fringe pattern and integrated PZT- piezo-MEMS actuator to showcase a preliminary result of on-chip beam shaping of the interference patterns.

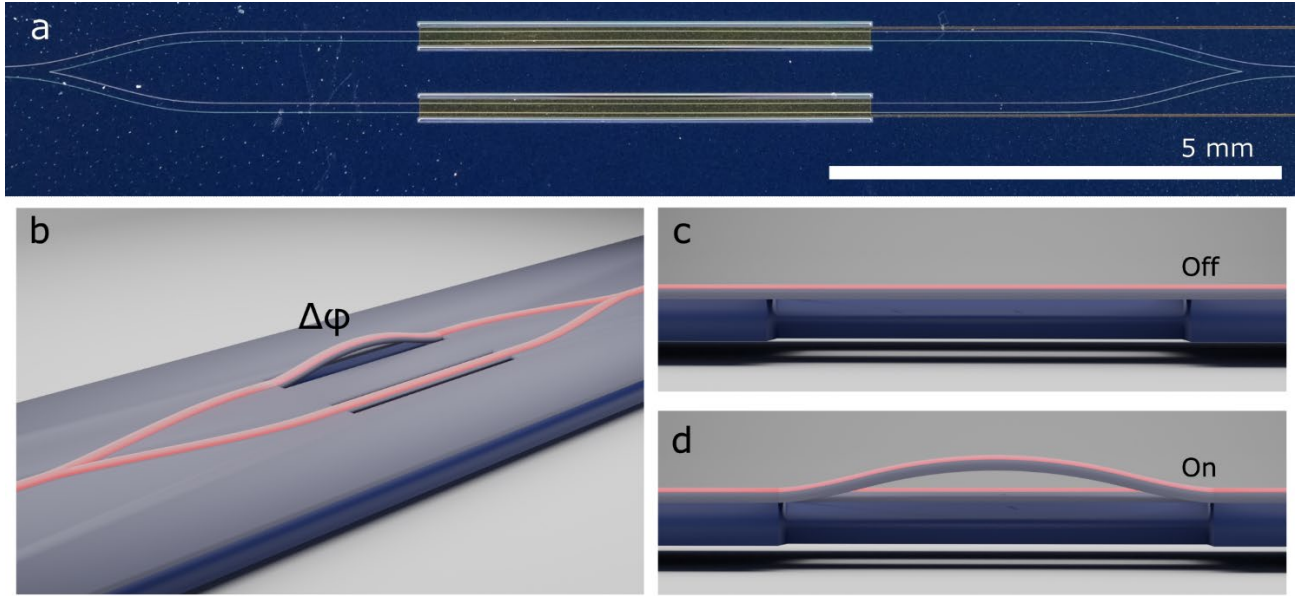

Figure 1: Concept and operating principle of the PZT MEMS actuator-based phase modulator. (a) Optical image of the fabricated chip showing the MZI waveguides and the integrated PZT stack. (b) Schematic illustration of the PZT MEMS actuator-based phase modulator. When an electrical field is applied to the PZT on the active arm, the free-standing suspended actuator undergoes physical deformation, while the reference arm remains unaffected. (c) Zoomed-in view of the reference arm (E = off) and (d) Zoomed-in view of the active arm under applied electrical field (E= on).

## Results and Discussion

The operating principle of the PZT-MEMS based phase modulator developed in this work is illustrated in Figure 1. The modulator is monolithically integrated into a MZI with single-mode waveguides to demonstrate phase shifting (Figure 1a). The MZI splits the optical signal into a reference arm and an active arm, which are later recombined to a single output waveguide. Sections of both arms are integrated with a PZT stack (shown in gold colored region in Figure 1a) and are subsequently released by selective front and back side etching to form suspended bridge-type actuators as illustrated in Figure 1b-d. To ensure symmetry and achieve high interference contrast, identical actuators are included on both arms. When an electric field is applied to the PZT stack on the active arm, in-plane contraction and out-of-plane expansion of the PZT layer induce bending of the suspended structure, including the PZT stack and underlying clamped waveguide (Figure 1d). This deformation changes the optical path length ($\Delta L$) of the active waveguide. In addition, stress-induced changes in the effective refractive index ($\Delta n_{\text{eff}}$) further contribute to the total phase shift ($\Delta \phi$) between the two interferometer arms). The reference arm remains straight (Figure 1c) since no field is applied. The phase difference can thus be expressed as:

$$\Delta \phi = \left(\frac{\partial \phi_{\text{act}}}{\partial E}\right) \Delta E_{\text{act}} - \left(\frac{\partial \phi_{\text{ref}}}{\partial E}\right) \Delta E_{\text{ref}} \qquad (1)$$

where $E_{\text{act}}$ and $E_{\text{ref}}$ represent the electric field applied across the PZT layer in the active and reference arms, respectively. The field-induced phase shift can be written as:

$$\frac{\partial \phi}{\partial E} = \frac{2\pi \ \Delta n_{\text{eff}} \ \Delta L \ (\Delta E)}{\lambda} \qquad (2)$$

λ is the wavelength in vacuum, L is the actuator length, and ΔE is the applied field step.

Beyond geometric deformation, the stress-optic contribution to phase shift was analytically estimated using Eq. 2, and data from Hassen *et al*. For a 10 mm long clamped PZT $Si_3N_4$ waveguide under 12 V, the stress-induced refractive index change ($\Delta n_{eff}$) is approximately $50 \times 10^{-6}$, corresponding to a $1.5\pi$ phase shift for 640 nm wavelength was reported[21]. Scaling this to a 3 mm long suspended waveguide using Eq. 2 predicts a stress-induced phase shift of $0.25\pi$ assuming the same $\Delta n_{eff}$ and wavelength (Figure S1b). In comparison, a pathlength difference of around 400 nm for 3 mm long suspended waveguide can result a $2\pi$ phase shift (Figure S1a), highlighting the significant footprint reduction enabled by the suspended geometry.

The mechanical stability of the bridge-type actuators critically depends on stress compensation across the multilayer stack. With the $Si_3N_4$ and surrounding oxide layer thicknesses fixed, we systematically varied the PZT thickness to identify the optimal stress balance. In-plane stress in the different thin film layers were measured and are listed in Table S1. Using an in-house analytical model, we predicted the PZT thickness that provides optimal overall stress equilibrium across the actuator stack, and the analysis indicates that a 1.0µm-thick-PLD-deposited PZT layer is close to optimal.

**Table 1:** Summary of the experimental measured values of the residual in-plane stress of the different thin films. These values were determined by measuring the film curvature and applying Stoney's formula to calculate the corresponding stress.

| **Thin films** | Si | Thermal $SiO_2$ | LPCVD $Si_3N_4$ | PECVD $SiO_2$ | Pt | PZT | Au |
|---|---|---|---|---|---|---|---|
| **Measured residual in-plane stress (MPa)** | N/A | -300 | 1100 | -120 | 1000 | 66 | 250 |
| **Thickness (µm)** | 8 | 2.5 | 0.140 | 1.5 | 0.1 | 1 | 0.25 |

The waveguides were simulated and designed using the commercial software Photon Design for single-mode operation at visible wavelength, λ = 635 nm. The $Si_3N_4$ waveguide core (thickness 135 nm, rib height of 5 nm) supports single-mode propagation for both TE and TM polarizations within 2 - 3 µm width, we have chosen 2.5 µm in this work. A rib geometry was selected over strip and slot designs due to reduced scattering loss, attributed to the minimal spatial overlap of the optical mode with the shallow-etched sidewalls. In addition, rib geometry allows fabrication simplicity, avoiding sub-micron widths required for single-mode strip guides at visible wavelengths[28]. The TM mode is used as it is more suitable for evanescent field-based sensing and imaging applications.

A summary of the fabrication process is outlined in Figure 2, and more details on the fabrication process can be found in supplementary section 2. Devices were fabricated on 150 mm double-sided-polished SOI wafers. After depositing the optical layers, 2.5 µm thermally grown $SiO_2$ and 140 nm LPCVD $Si_3N_4$, the waveguide were patterned using a two-step lithography and etching process to form a shallow (5 nm) rib waveguides with a 100 µm wide strip (as illustrated in Figure 2d). Following PECVD deposition of a 1.5 µm $SiO_2$ cladding (Figure 2e), the PZT stack (20 nm Ti/100 nm Pt bottom electrode, 1.0 µm PZT, 12 nm TiW/ 250 nm Au top electrode) was sequentially deposited by pulsed laser deposition (PLD) and patterned by a combination of wet and dry etching steps (Figure 2f).

Suspended bridge-type actuators were realized using a two-step process: front-side etching (Figure 2g) followed by back-side etching, both by combination of wet and dry (reactive ion etching) (Figure 2h). After the front-side etching the process wafer was bonded with a support wafer and through the backside etching fully suspended bridge-type actuators were produced. The processed wafers were diced and released from the support wafer. The dies were edge-polished to obtain high-quality coupling facets. Finally, the dies were mounted on PCBs and wire bonded for electrical actuation (as shown in Figure S3).

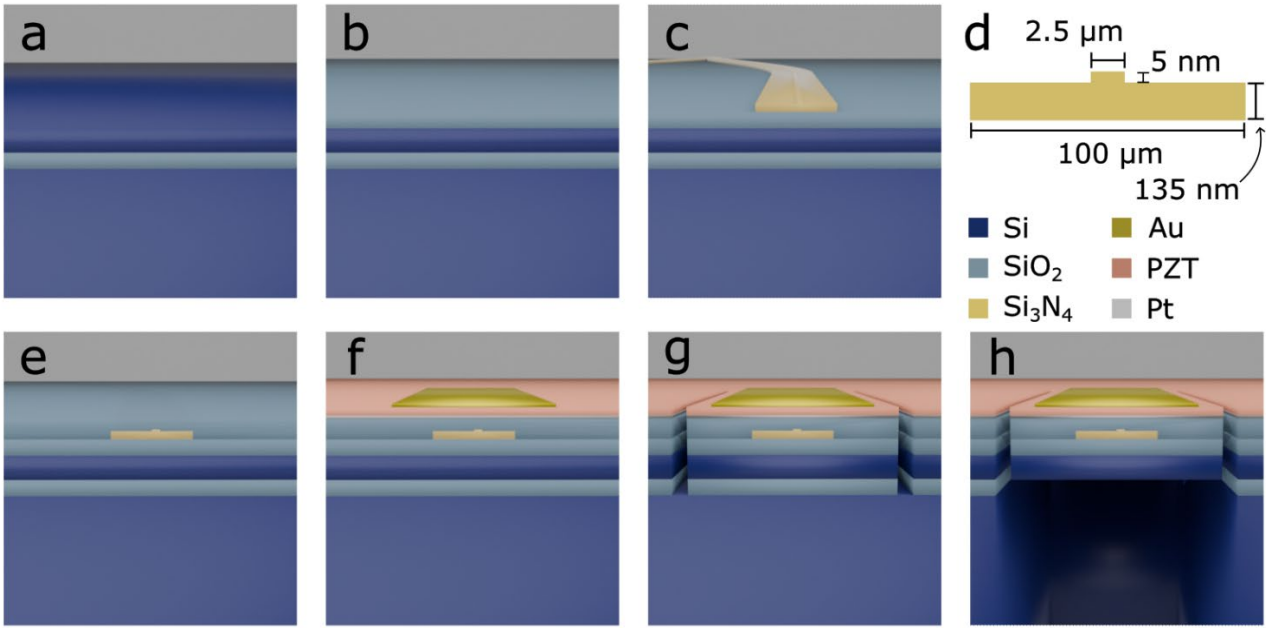

Figure 2: Main processing steps of the PZT-MEMS phase modulator. (a) SOI wafer, (b) thermally oxidized SOI wafer, (c) Si$_3$N$_4$ waveguide formation, (d) waveguide cross-sectional dimensions, the rib height of only 5 nm and the width of the rib waveguide is 2.5 µm, however the slab of 100 µm was etched for mechanical stability, (e) PECVD oxide cladding on top of the waveguides, (f) deposition and patterning of the PZT stack, (g) front-side etching, and (h) back-side etching to release the suspended actuator. The dimension of the suspended actuator in (g) and (h) is 200 µm wide and 3 mm long. This illustration is not drawn to scale.

The actuator displacement was characterized by using white-light interferometry (Figure 3a). In the unbiased state, the central displacement of suspended waveguide was approximately 25 µm upward, due to compressive residual stress in the PECVD and thermal oxide. The initial offset, however, did not inhibit tunability or cause significant losses upon PZT-actuation. Although analytical optimization was performed during designing of the actuator to minimize residual stress, some deformation remained after fabrication. At voltages from 0 to 10V, the 3 mm long active arm exhibited a progressive vertical displacement (Figure 3c), corresponding to a maximum 175 nm differential of elongation (Figure 3d). This led to an experimentally measured phase shift of $1.42\pi$ (Figure 4), closely matching the analytical predication of $1.4\pi$ (Figure S1). The measured/simulated path along the actuator were integrated to numerically calculate the elongation.

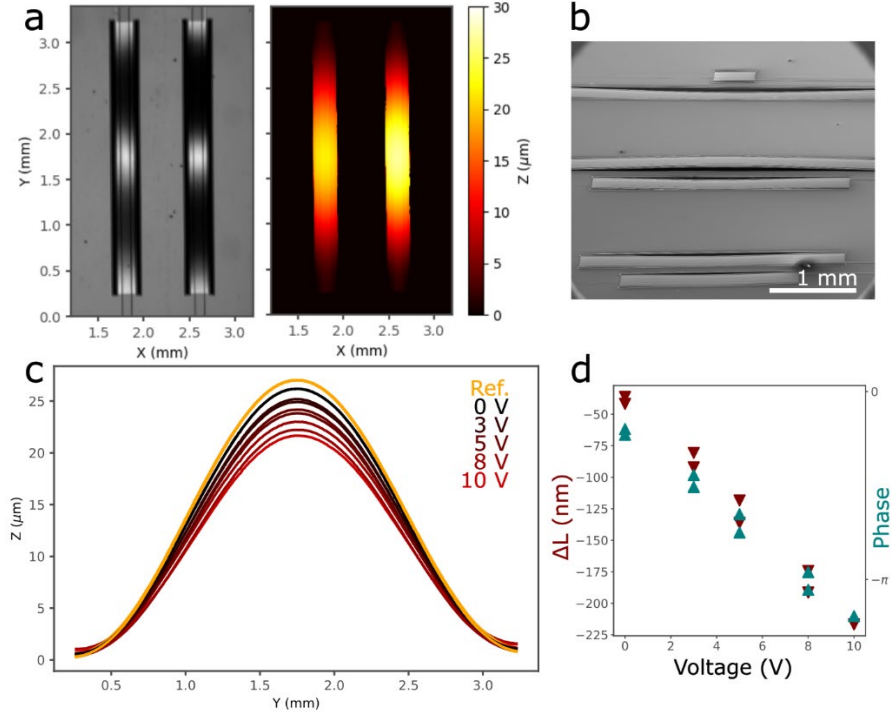

Figure 3: White-light interferometer and SEM characterizations. (a) Optical/white-light interferometer images of the MZI structures. (b) SEM image showing the MEMS from an oblique angle. (c) Line profiles along the actuators at different applied voltages. (d) Extracted displacement values and corresponding theoretical phase shifts.

The optical characterizations of our platform were performed using the optical setup illustrated in Figure S3. The details of the experimental setup are described in supplementary section 3. The propagation loss of the waveguide, including the suspended region, was measured to be 0.75 dB/cm for TM-polarized light at 635 nm (measurement method described in supplementary section 4), confirming that the suspended region of the waveguide does not add high losses. This is mainly due to the smooth transition of the guided mode from the clamped to the suspended region. The guide mode remains well confined in the $Si_3N_4$ layer by the underlying thermal oxide and device layers, and also a minuscule waveguide bending angle introduced in the suspended region.

The experimental phase shift measurements were done for two actuators with length of 3 mm (device A) and 5 mm (device B). As the applied voltage varies, the output intensity of the MZI waveguide changes. The phase shift ($\Delta\phi_s$) was extracted from the recorded output intensity using the relation:

$$\Delta\phi_s(t) = \cos^{-1} \frac{I_T(t) - I_{act} - I_{ref}}{2\sqrt{I_{act} I_{ref}}} \qquad (4)$$

Here, $I_{act}$ and $I_{ref}$ are the intensity of the active and reference arms, determined from the measured maximum ($I_{max}$) and minimum ($I_{min}$) the output intensity ($I_T$): $I_{act} + I_{ref} = (I_{max} + I_{min})/2$ and $2\sqrt{I_{act} I_{ref}} = (I_{max} - I_{min})/2$. When the phase shift exceeds $\pi$, phase unwrapping is applied to ensure continuity.

Figure 4 shows the measured phase shift and time response for Device A as a function of applied voltage. Figure 4a-c shows the applied voltage, the output intensity and the extracted phase (using Eq. 4), respectively. For device A, a $\pi$ phase shift was achieved at 8.25 V, and a 1.45$\pi$ phase shift at 10V (Figure 4d), this is in a close agreement with the phase shift calculated from white-light interferometer displacement measurements Figure 3d. Device A has a rise time ($t_{10-90}$) of around 5 ms (Figure 4e), corresponding to 200 Hz. The measured phase shift for Device B is shown in Figure S5. For the Device B, a π phase shift was achieved at 3 V, and 2.5 π at 10 V, with the rise time of around 2.6 ms.

Nonlinearity in the tuning and the upward and downward voltage sweep (Figure 4d and S5e), stems from the ferroelectric nature of the PZT thin film[29]. Figure S6 shows the measured polarization (P), capacitance (C), loss tangent (tan(δ)) and relative permittivity for electrode dimensions of 200 µm by 3 mm and a PZT thickness of 1µm using an aixACCT TF3000 ferroelectric tester. P = 18.4 µC/cm$^2$, C = 6.5 nF, $\varepsilon_r$ = 1233, and tan(δ) = 0.05, are indicates typical high-quality PZT thin films[27]. Using these values, the stored energy at 8.5V (corresponding to a π phase shift) is E=$\frac{CV^2}{2}$ ~ 235 nJ. The energy dissipated per cycle due to dielectric losses $E_{loss}$ ~ E*tan(δ), $E_{loss}$ ~ 11.75 nJ per cycle. For quasi-DC operation, (1Hz), the corresponding power dissipation is 11.75 nW per π phase shift, increasing to 2.53 µW at 215 Hz. For quasi-static applications, a closed loop operation is needed to compensate for the hysteresis, e.g. through capacitive, piezoresistive or optical position read-out of the bridge-structure. For the current harmonic application, however, hysteresis is not an issue due to dynamic actuation. i.e., the same path is traversed repeatedly, and control can be ensured through voltage table look-up.

In the present design, the actuator and electrode width is 200 µm (Fig 2g-h), however, the optical waveguide is only 2.5 µm wide (Fig. 2d). Thus, in future designs, the actuator and electrode width can be substantially reduced without compromising performance. For more compact actuator with electrode dimensions of 20 µm by 3mm, capacitance scales proportionally and the reducing the expected power consumption at 500Hz to approximately 0.34 µW.

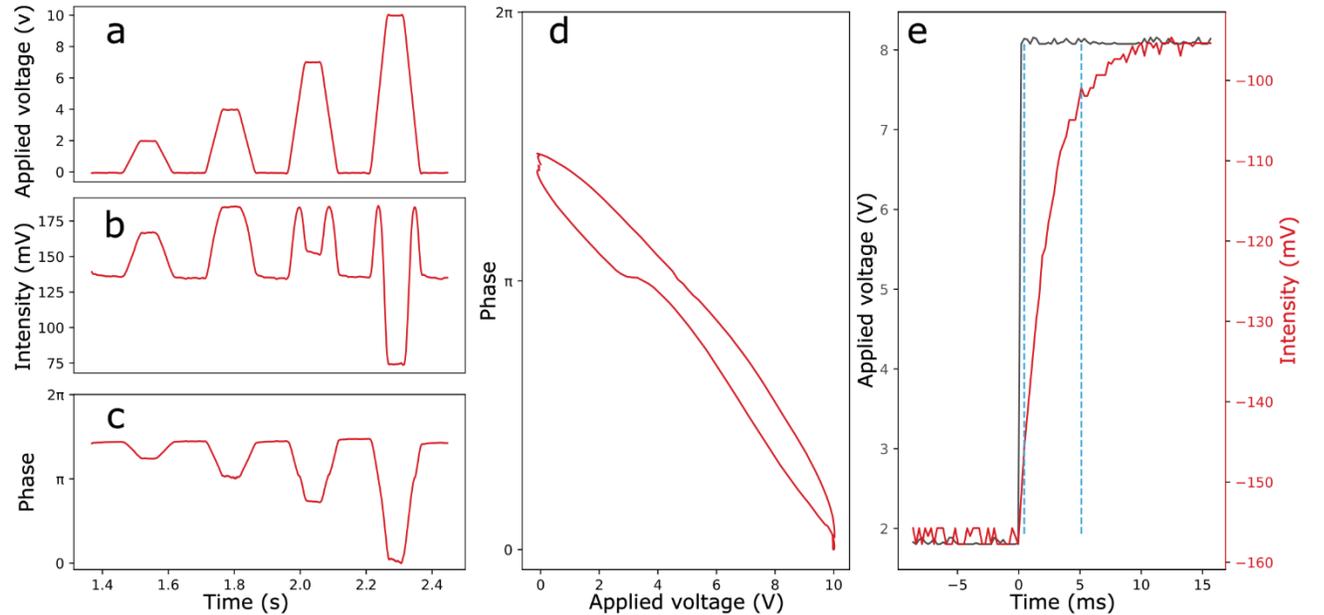

Figure 4: Experimental characterization of the phase shift in Device A. (a) Sequential voltage steps of 0, 2, 4, 8, and 10 V were applied, allowing the system to stabilize at each step before returning to 0 V. (b) Measured output intensity of MZI as a function of time, showing clear correspondence with the applied voltage pattern. (c) Calculated and unwrapped phase shift extracted from the output intensity. (d) Extracted phase as a function of applied voltage, showing a phase shift of π at 8.25 V and 1.45 π at 10 V (e) Measured rise time, $t_{10-90}$ = 4.7 ms, indicated by the dashed blue lines.

Finally, we present preliminary results showcasing on-chip beam shaping[30] of evanescent interference fringes using the PZT-MEMS phase modulator. An experimental setup similar to that used for on-chip structured illumination microscopy (SIM) was used[6]. This technique allows us to create intensity fringes within an area on which biological tissue is placed. By phase modulating using the active bridges, the fringes can be moved and three phase shifted images can be acquired to obtain two times gain in the optical resolution as supported by the SIM algorithms. The interferometer design and the beam shaping results are shown in Figure 5. A single mode rib waveguide was spilt into two arms and recombined at the imaging area (shown by circle in Figure 5a and 5d). The single mode was adiabatically tapered to 50 μm to form interference fringes over large area, as shown in Figure 5a and 5d.

To image the interference fringes we coated fluorophores onto of the waveguide surface as explained next. Poly-L-lysine (PLL) was pipetted inside the PDMS chamber enclosing the imaging area. After 20 minutes, the PLL was rehydrated using distilled water and washed away. A few microliters of cell mask orange fluorophore were deposited on the imaging area and sealed with a #1.5 cover slip for imaging. The deconvolved interference fringe image in Figure 5c was acquired with a 561 nm excitation laser, a suitable filter set (585 nm long-pass and 630/75 nm bandpass) and imaged using a 60X 0.9 N.A. objective lens. Figure 5b shows the fringe patterns passing over a scattering point shown in circle in Figure 5c as a voltage from 0 to 10V was applied to one of the 3 mm long actuator, repeatedly three times. The fringe contrast is relatively poor, which may be attributed to one or more of the following factors: poor quality of the waveguide surface, accumulation of dirt and unwanted scratches on the surface during sample preparation, non-uniform coating of the fluorophore itself and background light in the slab region. Nevertheless, the preliminary results show the utility of an ultra-low power active PICs based on $Si_3N_4$ platform for on-chip beam engineering at visible spectrum.

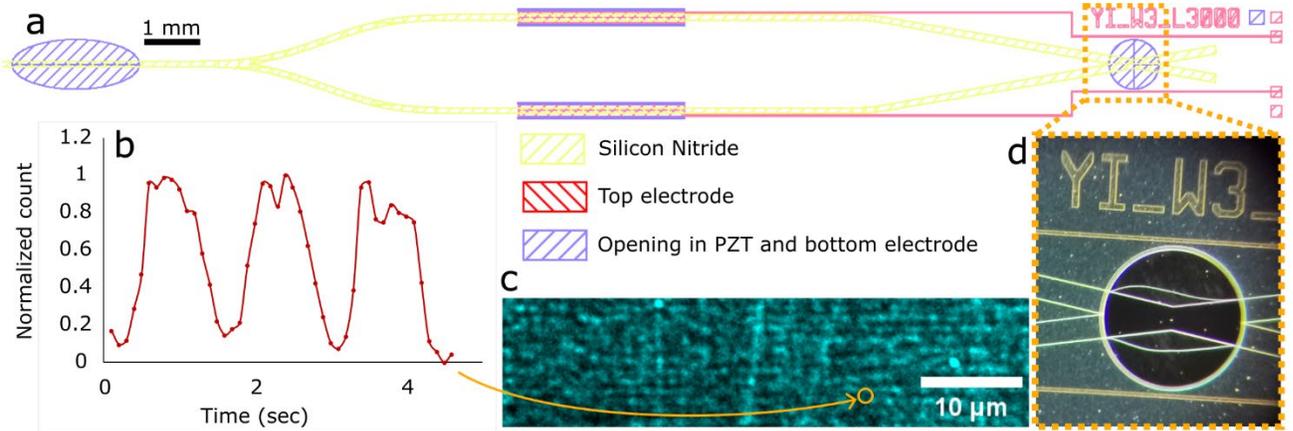

Figure 5: PZT-MEMS based phase modulator for on-chip beam shaping. (a) The interferometer design, where two guided beams interfere to form evanescent field fringe pattern, with a 3 mm long actuator integrated in both arms. (b) Fringe displacement as a function time (indirectly measuring fringe displacement as a function of voltage 0 to 10V applied three times) measured in a small area shown in (c) and highlighted by the circle. (c) Fluorescence microscopy image of the generated fringe patterns obtained on top of waveguide surface (single-timepoint). (d) Bight field optical image of the beam interference region corresponding to the device layout.

In this work, we have demonstrated a PZT-MEMS-based phase modulator for $Si_3N_4$ photonic waveguide operating in the visible spectrum. By combining a high-piezoelectric-coefficient PZT stack with a $Si_3N_4$ waveguide and implementing a suspended bridge-type actuator, we achieve mechanically induced phase modulation with high efficiency and a compact footprint. When incorporated into MZI, the proposed devices exhibit large phase shifts, up to $1.4\pi$ and $2.5\pi$ for 3 mm and 5 mm actuators at 10 V, respectively, corresponding to a scalability metric ($V\pi \cdot L$) of 2.25 V·cm. The modulator operates with ultralow power consumption (~5.4 nW per π phase shift for quasi-DC operation and 2.71μW at 500Hz) and maintain low optical loss efficiency ($V\pi \cdot L \cdot \alpha$) of 0.75 dB/cm at 635 nm, confirming the effectiveness of the suspended actuator design. Compared to existing visible-wavelength $Si_3N_4$ modulators, our approach provides an order-of-magnitude improvement in scalability metric over stress-optic PZT devices and a 50% enhancement relative to AlN-based piezoMEMS modulators, demonstrating its scalability and performance advantage.

Future efforts will focus on three main directions: (1) increasing modulation speed by reducing the actuator length, acknowledging that this will high power consumption to achieve the required phase shift; (2) reducing power consumption by using narrower actuator and electrodes designs; and (3) enhancing mechanical deformation by decreasing the actuator thickness for example, by thinning the device layer beneath the actuator stack and waveguide layer. The PZT-MEMS phase modulator concept is broadly compatible with PIC platforms, as it relies on integrating a piezoelectric thin film on a cladded waveguide. This make it particularly attractive for waveguides platforms lacking intrinsic thermo-optic, electro-optic, or piezoelectric effects, such as aluminium oxide ($Al_2O_3$)-based photonics waveguide[31] which is gaining popularity owing to its broad-spectrum transmission window from ultraviolet to mid-IR.

Overall, these results position PZT-MEMS phase modulator on $Si_3N_4$ ($Al_2O_3$) waveguide as a promising route for low-power, compact, and high-performance active modulation in visible and ultraviolet spectrum photonics, paving the way for scalable on-chip beam steering, programmable photonics, bioimaging, and quantum technologies. The combination of ultralow power consumption and compact device footprint renders PZT-MEMS–based phase modulator particularly appealing for applications that demand dense phase-modulator integration, for example in optical phased arrays.

## Acknowledgement

The author would like to thank Andreas Vogl, Jesil Jose, Karolina Barbara Milenko-Kuszewska, Aina K. Herbjønrød, Runar Dahl-Hansen (SINTEF Digital, Oslo, Norway), and Frode Tyholdt (Sonair, Oslo, Norway) for their valuable feedback and helpful suggestions. This research was funded by the Research Council of Norway, project numbers 194068/F40 and 352764.

## Supporting Information Available

Further information on fabrication and characterization can be found on-line: